# Calculating the contribution of different binding modes to Quinacrine–DNA complex formation from polarized fluorescence data

I.M. Voloshin, O.A. Ryazanova, V.A. Karachevtsev and V.N. Zozulya

*Department of Molecular Biophysics, B. Verkin Institute for Low Temperature Physics and Engineering of NAS of Ukraine, 47 Lenin ave., 61103, Kharkiv, e-mail: voloshin@ilt.kharkov.ua*

Binding of acridine derivative quinacrine (QA) to chicken erythrocyte DNA was studied by methods of absorption and polarized fluorescent spectroscopy. Measurements were carried out in aqueous buffered solutions (pH 6.9) of different dye concentrations ([QA] = $10^{-6} \div 10^{-4}$ M) and ionic strengths ([Na$^+$] = $10^{-3} \div 0.15$ M) in a wide range of phosphate-to-dye molar ratios (*P/D*). It is established that the minimum of fluorescent titration curve plotted as relative fluorescence intensity vs *P/D* is conditioned by the competition between the two types of QA binding to DNA which posses by different emission parameters: (i) intercalative one dominating under high P/D values, and (ii) outside electrostatic binding dominating under low P/D values, which is accompanied by the formation of non-fluorescent dye associates on the DNA backbone. Absorption and fluorescent characteristics of complexes formed were determined. The method of calculation of different binding modes contribution to the complex formation depending on P/D value is presented. It was shown that the size of binding site measured as the number of DNA base pairs per one QA molecule bound in the case of the electrostatic interaction is 8 times less than that for the intercalative one that determines the competitive ability of the outside binding against the stronger intercalative binding mode.

**Key words:** quinacrine, DNA, outside binding, intercalation, cooperative binding, fluorescence, absorption.

## I. INTRODUCTION

Study of interaction between different organic dyes and intracellular biopolymers is one of major scientific approaches which give the opportunity to obtain information about molecular mechanisms of biological system activities. In particular, the dyes give the possibility to visualize the intracellular structures. So, after the works of Caspesson et al. [1-3] devoted to staining of fixed chromosomes, acridine derivative called quinacrine (QA – quinacrine, mepacrine, atabrine – Fig.1) become widely applied in cytology. This dye fluoresces intensively in yellow-green range of spectra, that gives the possibility to obtain the specific staining pattern (characteristic transverse bands that appear on chromosomes), that was successfully used under the chromosome analysis. High biological activity provides wide range of its application in biology and medicine as antiprotozoal, antirheumatic, antihelmintic agent. Quinacrine was first synthesized by Bayer (Germany) in 1931. During the II World War it has been widely used for prevention and therapy of malaria. In the present time QA is widely used for therapy of lambliasis, diphyllobothriasis, leishmaniasis, taeniasis, lupus erythematosus et al. [4, 5]. It was shown that QA bound to prion peptides blocks their transformation to pathogenic form that makes it promise candidate for design of drugs against prion encephalopathy [6]. Such a wide range of QA action causes a great scientific interest and the need to clarify the molecular mechanisms of complex formation between them and biopolymers, as well as to account their structural features. The dye represents dication compound with planar aromatic tricyclic structure, therefore its interaction with charged nucleic acids (NA) and peptides is of special interest. Being the key participants of biological processes, these biopolymers undergo the structural transformations as a result of the complex formation, which can affect their functioning. For example, interaction of DNA with acridine derivatives leads to the frameshift mutation, photocleavage and another effect [7-10].

It is well known that under the interaction of acridine derivatives with DNA, together with strong intercalation binding, weaker external binding takes place which is conditioned by electrostatic coupling of positively charged chromophores with negatively charged phosphate groups of DNA backbone [11, 12]. Its contribution to the complex formation can be significant even at physiological ionic strength of solution and high values of molar phosphate-to-dye ratios (*P/D*) [13, 14]. However, in spite of great number of works

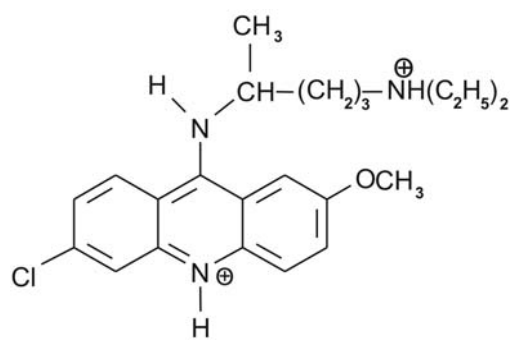

devoted to study of the dye–NA interaction, up to now no any reliable data about quantitative ratio between contributions of these binding modes that can be conditioned by small number of experimental techniques suitable to separate them efficiently. Analysis of literature available shows that main attention was paid to the intercalation binding of acridines to NA at high

FIG. 1. Molecular structure of quinacrine dye (QA).

*P/D* ratios [15]. The contribution of the external electrostatic binding was neglected, considering it as unsubstantial at physiological ionic conditions. This could leads to some error in the evaluation of binding cooperativity. External binding have also to be taken into account when using fluorescent probes for investigation of intracellular processes.

Polarized fluorescent spectroscopy is powerful experimental technique which can be used to estimate quantitatively contribution of different binding modes to dye–NA complex formation since it is very sensitive to any changes in environment properties and chromophore mobility. However the data about changes in fluorescence polarization degree of organic dyes during its binding to NA are very limited. In particular, no data available on the change in fluorescence polarization degree when QA binds to DNA in the wide range of concentrations, different solution ionic strength and *P/D* ratios.

In the present work binding of quinacrine to DNA have been studied at different experimental conditions (in the wide range of phosphate-to-dye molar ratios, *P/D*; at different values of dye concentrations ($10^{-6} \div 10^{-4}$ M) and solution ionic strength (0.0012, 0.01 and 0.15 M Na$^+$)) using techniques of absorption and polarized fluorescent spectroscopy to estimate quantitatively the contributions of intercalative and external electrostatic binding to complex formation.

## II. EXPERIMENTAL

The quinacrine dihydrochloride (QA, Fig. 1) was purchased from Serva (Heidelberg, Germany). High-polymer DNA sodium salt from chicken erythrocyte (containing 41% of GC base pairs) was obtained from Reanal (Hungary).

For all experiments 1 mM sodium cacodylate buffer, pH 6.9, containing 0.1 mM Na$_2$EDTA and prepared in fresh deionized distilled water to which NaCl (Sigma Chemical Co.) was added to give concentrations of 0.01 and 0.15 M Na$^+$. Buffers were ultra filtered through nitrocellulose filters (Millipore-Q system, USA) with a pore diameter of 0.22 microns.

The concentrations of the dye and DNA were determined spectrophotometrically using the molar extinction coefficient of $\varepsilon_{260} = 6600$M$^{-1}$cm$^{-1}$ for DNA and $\varepsilon_{424} = 9750$ M$^{-1}$см$^{-1}$ for QA [16]. Samples for spectral measurements were prepared by mixing the stock solutions of known concentration in predetermined volume proportions.

Electronic absorption spectra were measured on a SPECORD UV-VIS spectrophotometer (VEB Carl Zeiss, Jena). Measurements of steady-state fluorescence intensity were carried out by the method of photon counting with a laboratory spectrofluorimeter based on double monochromator DFS-12 (LOMO, Russia). Fluorescence excitation was performed by linearly-polarized beam of He-Cd laser LPM-11 ($\lambda = 441.6$ nm), which power was stabilized during the experiment using hand-made set-up described in [17]. The fluorescence intensity was registered at the right angle to the incident beam. Ahrens prisms were used to polarize linearly the exciting beam as well as to analyze the fluorescence polarization. The spectrofluorimeter was equipped with a quartz depolarizing optical wedge to exclude the monochromator polarization-dependent response. When measuring the fluorescence intensity, the pulses from photomultiplier tube were accumulated during 10 s for each data point and measurements were repeated five times, at that the measurements error was about 0.5%. Also the correction was made to the absorption of the laser beam in the solution layer of the front wall to the cell center. Fluorescence spectra were corrected on the spectral sensitivity of the spectrofluorimeter. Experimental set-up and the measurement procedure were described earlier [18]. The total fluorescence intensity, $I$, its polarization degree, $p$, and anisotropy, $\mu$, were calculated using formulas [19]:

$$I = I_{II} + 2I_{\perp} \tag{1}$$

$$p = \frac{I_{II} - I_{\perp}}{I_{II} + I_{\perp}} \tag{2}$$

$$\mu = \frac{I_{II} - I_{\perp}}{I} = \frac{2p}{3 - p} \tag{3}$$

where $I_{II}$ and $I_{\perp}$ - are measured intensities of the emitted light, which are polarized parallel and perpendicular to the polarization direction of the exciting light beam, respectively.

The binding of quinacrine to DNA was followed by changes in parameters of the dye fluorescence under titration experiments at several fixed values of QA concentration being in the range of $10^{-6} \div 10^{-4}$ M. The QA solution was added with QA–DNA complex containing the same concentration of the dye that gives the possibility to obtain the sample of required $P/D$ ratio without changing the concentration of the dye.

Fluorescence intensity and polarization degree of complexes were registered at the wavelength corresponding to the maximum of free QA, $\lambda_{obs} = 510$ nm.

All measurements were carried out in 1 cm quartz cell at room temperature from 22 to 24 °C.

### III. RESULTS AND DISCUSSION

Visible electronic absorption and fluorescence spectra of quinacrine in a free state and bound to DNA are depicted in Fig. 2. Absorption spectrum of the free dye represents the superposition of two intense bands with maxima at 424 and 445 nm and less intensive shoulder. The fluorescence spectrum is a broad intense unstructured band centered at 510 nm. Fluorescence polarization degree, *p*, amounts to 0.035.

The results of the titration of QA solution with QA–DNA complex are presented in Figs. 3(a,b) as *P/D* dependent changes in the relative fluorescence intensity, $I/I_0$ (Fig. 3a), and polarization degree, *p* (Fig. 3b). Here $I_0$ and $I$ are fluorescence intensities of the free and bound dye correspondingly, measured at $\lambda = 510$ nm. The curves illustrate dependence of QA binding to DNA on the solution ionic strength at the low dye concentrations ($C_{QA} = 10^{-6}$ M). It is clearly seen that the binding results in the strong changes in the QA absorption and fluorescent characteristics, especially at *P/D* < 30 where two parts of the titration curves can be identified. So in the solutions of low ionic strength, $[Na^+] = 1.2$ mM, the initial part of titration curve ($P/D = 0 \div 4$) is linear, that is typical for outside electrostatic binding of cationic dyes to polyanions [20-22].

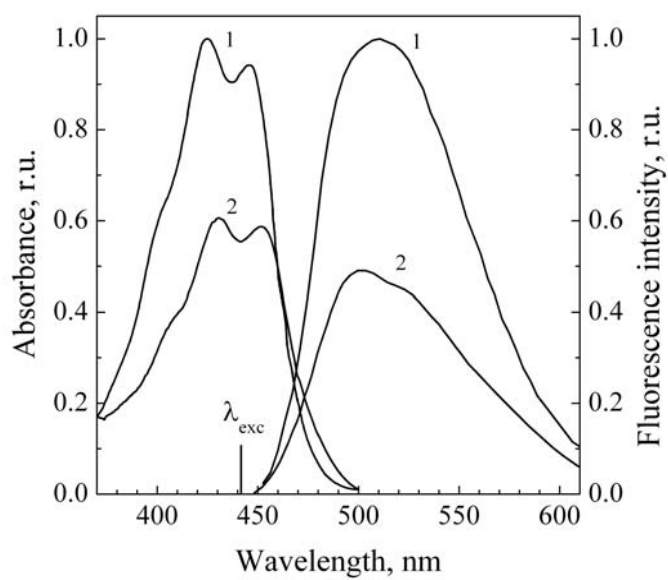

FIG. 2. Normalized absorption (left) and fluorescence spectra (right) of free quinacrine (1) and its complex with DNA registered at $P/D = 100$ (2) in aqueous buffered solution at room temperature.

Increase in solution ionic strength reduces the dye binding to DNA due to the competition between QA molecules and sodium ions, that results in decrease of the depth of fluorescence titration curve minimum for the sample with $[Na^+] = 0.01$ M ($I/I_0 = 30$ %) and its total disappearance in sample of near physiological ionic strength, $[Na^+] = 0.15$ M. In the last case the *P/D* dependent rise of fluorescent polarization degree is substantially slighter in comparison with that in solution of low ionic strength (Fig. 3b). This evidences that the sodium ions weaken not only outside electrostatic binding of QA to DNA, but the intercalative one also. However fluorescence polarization degree also reaches the constant level of *p* = 0.3 at $P/D \approx 200$ for the sample containing 0.01 M of $Na^+$, and at $P/D \approx 1000$ for that with $[Na^+] = 0.15$ M (not shown).

In Fig. 4 we can see the fluorescent titration curves registered in solutions of low and high ionic strengths at higher concentration of the dye, $C_{QA} = 10^{-4}$ M. For these samples stronger fluorescent quenching were observed as compared with previous system ($C_{QA} = 10^{-6}$ M, Fig 3a): in the curve minimum at $P/D = 4$ the emission is only 9 % from initial, since the fraction of the dye externally bound to DNA is proportional to its concentration.

More visually the dependence of QA to DNA binding on dye concentration are demonstrated in Fig. 5, where QA–DNA fluorescent titration curves are represented for the three samples at the following dye concentrations, $C_{QA} = 10^{-6}$; $2\cdot10^{-5}$; $10^{-4}$ M, and low Na$^+$ content (1.2 mM). From the figure it is seen, that that growth of QA concentration strengthens appreciably quenching of its emission that results in a deepening of a minimum on the titration curve. However, practically total fluorescence quenching have not observed, as it was registered earlier for QA complex with inorganic polyphosphate (see dashed line in Fig. 5 [27]), where only pure electrostatic binding occurred, since even at $P/D = 4$ nonzero contribution of intercalation binding mode to QA–DNA complex formation takes place.

Simultaneous measurements of fluorescence intensity and anisotropy of complexes *vs* $P/D$ ratio allow to separate the contributions of outside and intercalative binding to QA – DNA complex formation, as well as to determine the fraction of free and bound dye molecules in the samples, since these parameters are substantially different.

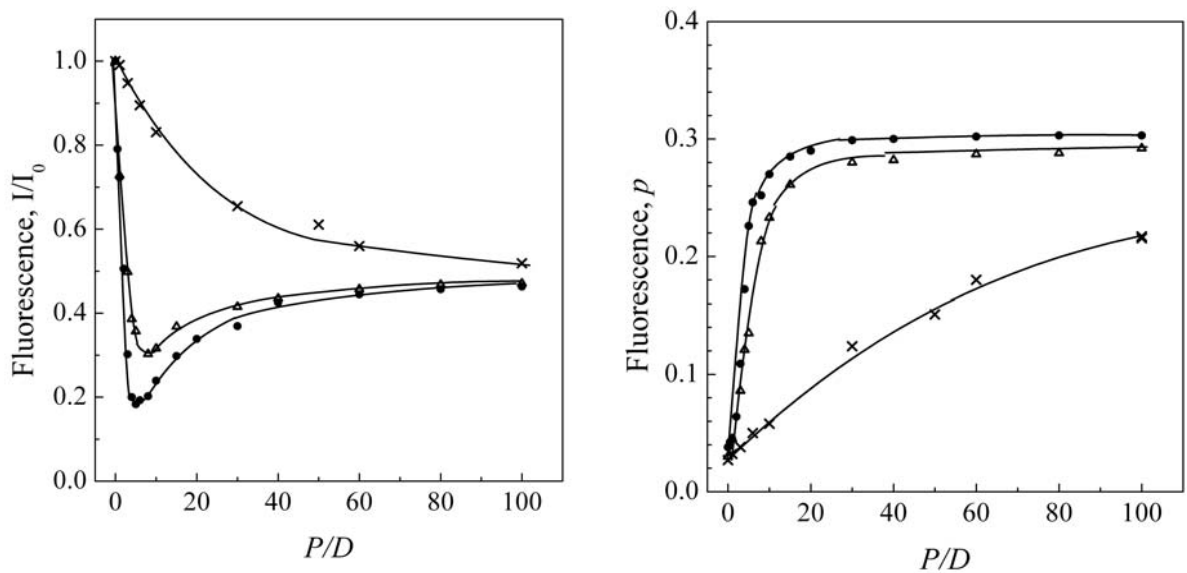

FIG. 3. Dependence of QA normalized fluorescence intensity (left) and polarization degree (right) on molar DNA-to-dye ratio ($P/D$) obtained in the solutions of different ionic strength: (●) – 1.2 mM Na$^+$, (△) – 0.01 M Na$^+$, (✗) – 0.15 M Na$^+$, $C_{QA} = 10^{-6}$ M; $\lambda_{exc} = 441.6$ nm; $\lambda_{obs} = 510$ nm.

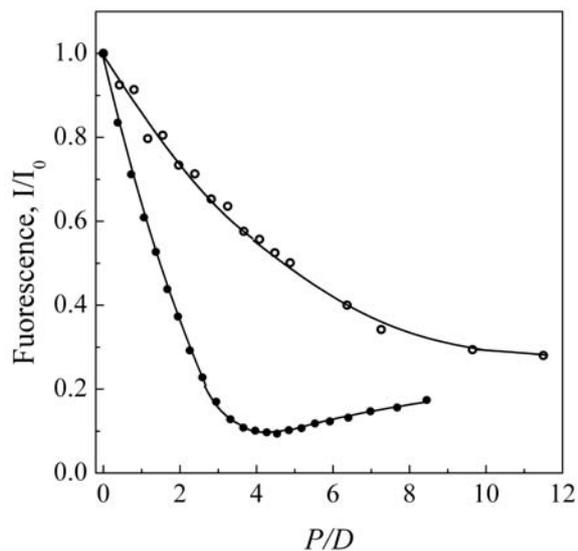

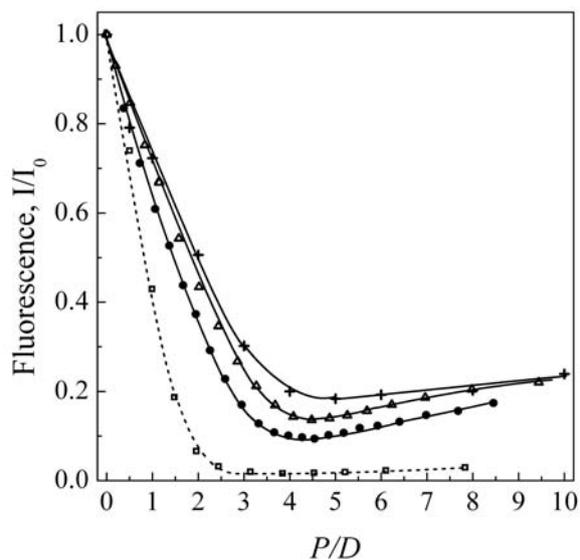

FIG. 4. Titration curves plotted as normalized fluorescence intensity *vs* *P/D* for QA–DNA complexes, $C_{QA} = 10^{-4}$ M, in solutions of different ionic strength: (●) – 1.2 mM Na$^+$, (○) – 0.1 M Na$^+$; $\lambda_{exc} = 441.6$ nm; $\lambda_{obs} = 510$ nm.

FIG. 5. Titration curves plotted as normalized fluorescence intensity *vs* *P/D* for QA–DNA complexes at following QA concentrations: (●) – $10^{-4}$ M, (△) – $2 \cdot 10^{-5}$ M, (+) – $10^{-6}$ M, in solutions containing 1.2 mM Na$^+$; $\lambda_{exc} = 441.6$ nm; $\lambda_{obs} = 510$ nm. Dashed line corresponds to quinacrine–polyphosphate complex, $C_{QA} = 5 \cdot 10^{-5}$ M.

For this purpose, the system of equations were used [28,29]:

$$\sum_{i=0}^{2} \mu_i I_i f_i = \mu I \qquad (4)$$

$$\sum_{i=0}^{2} I_i f_i = I \qquad (5)$$

$$\sum_{i=0}^{2} f_i = 1 , \qquad (6)$$

where $f_i$ - fraction, $I_i$ – fluorescence intensity and $\mu_i$ – fluorescence anisotropy for the dye in $i$-state; here $i = 0$ corresponds to the free QA, $i = 1$ – to the dye intercalated into DNA and $i = 2$ – to externally bound molecules. From the data obtained the plots of $f_i$ vs *P/D* were constructed (Fig. 6). The curves presented in this figure relates to QA–DNA samples containing $2 \cdot 10^{-5}$ M of quinacrine and 1.2 mM Na$^+$. They illustrate the contributions of two QA–DNA binding modes at different ratios of the dye and polymer in solution. So, intercalative binding mode dominates at high *P/D* values, whereas at low ones an external electrostatic binding prevails. With decreasing *P/D*, the population density of intercalation binding sites is reduced, that leads to the redistribution of QA molecules between two bound states in favor of external binding. For example, at $P/D = 50$ approximately 91.2 % of quinacrine molecules are intercalated to double helix, 8,3 % involved to external electrostatic binding, and only 0.5 % does not participate in the complex formation. In the minimum of fluorescent titration curve at $P/D = 4$ the major part of dye molecules (78.5 %) are included in the complexes of second type (external binding), and only 14 % of the molecules are intercalated into double helix; 7.5 % of the dye remains unbound.

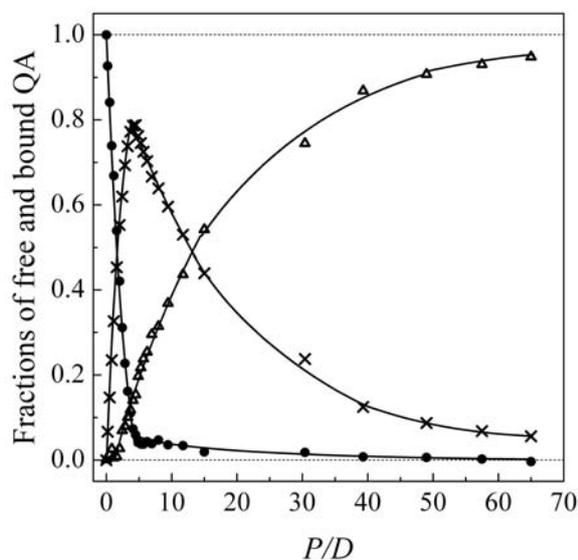 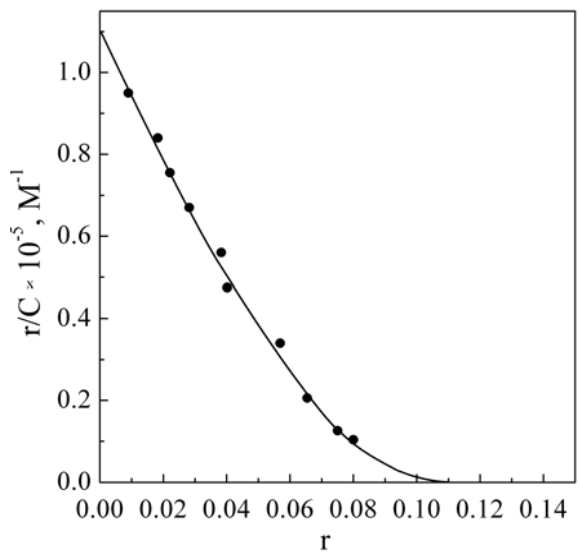

FIG. 6. Fractions of QA in a free state (●) and bound to DNA: (△) – intercalation, (✗) – external binding. Data were calculated for QA-DNA complexes ($C_{QA} = 2 \cdot 10^{-5}$ M) in the solution containing 0.0012 M $Na^+$.

FIG. 7. Scatchard plot constructed for QA bound to DNA in the solution containing 0.15 M $Na^+$, $C_{AX} = 10^{-6}$ M. Solid line corresponds to the curve calculated from McGhee and von Hippel equation (7) at $K = 1.1 \cdot 10^5$ $M^{-1}$ and m = 8.

Parameters of QA intercalative binding to DNA were determined from Scatchard plot (Fig. 7) constructed for the sample with low dye concentration, [QA] = $10^{-6}$ M, and high content of sodium ions, 0.15 M $Na^+$, where external electrostatic binding can be neglected. This binding type is characterized by non-linear adsorption isoterm, which is well described by theoretical dependence, obtained for non-cooperative binding of the ligands to homopolymer [30]:

$$\frac{r}{C} = K \frac{(1 - mr)^m}{[1 - (m-1)r]^{m-1}} \qquad (7),$$

where $K$ is apparent binding constant, $m$ is a size of binding site, equal to the number of base pairs per one dye molecule, $r$ – number of bound dye molecules per one base pair, $C$ – concentration of free dye. The best fit of experimental points was obtained using the next parameter in the equation (7): $m = 8$, $K = 1.1 \cdot 10^5$ $M^{-1}$ (Fig. 7). The result means that the saturation of intercalation binding occurs at one dye molecule per 8 base pairs of DNA. The obtained $m$ value of is in good agreement with the data [31]. Under reduction of solution ionic strength to $[Na^+] = 1.2$ mM the binding constant $K$ characterizing the intercalation increases to approximately $2 \cdot 10^9 \, M^{-1}$ [16].

The external electrostatic binding is characterized by substantially smaller binding sites, m = 1, and apparent binding constant which is three orders of magnitude less than that for the intercalation. It was estimated using model system of quinacrine – inorganic polyphosphate where only the pure electrostatic binding of QA to DNA occurs. In solution with low $Na^+$ content, 1.2 mM, the apparent binding constant was found to be equal $K = 1{,}7 \cdot 10^6 \, M^{-1}$ [27]. The described above transition of the dye from the intercalated to externally bound state can be explained using McGhee and von Hippel theory [30, 32], videlicet, under the binding saturation the total increment of free energy can be greater in the case of less strong interaction at the cost of higher density of the lattice binding sites.

## IV. CONCLUSIONS

Quinacrine binding to DNA is realized via two competing mechanisms: (i) intercalation of the dye between the biopolymer base pairs and (ii) external electrostatic binding of the dye to DNA polyanionic backbone with $\pi$-$\pi$ stacking of its chromophores. Simultaneous measurements of fluorescence intensity and polarisation degree give the possibilty not only determine the fraction of free and bound dye molecules, but quantitatively separate the contributions of these two binding modes. It was shown that in the case of outside electrostatical binding the size of binding site (measured in the DNA base pairs per one quinacrine molecule) is 8 times less than that for intercalation, that allows him to compete successfully. Upon the saturation of the QA binding an external type of complex formation can be dominant.